\documentclass[useAMS,usenatbib]{mn2e}

\usepackage{amsmath}

\usepackage{graphicx}

\title[A Spallation Model for $^{44}$Ti production in Core Collapse Supernovae]{A Spallation Model for $^{44}$Ti production in Core Collapse Supernovae}
\author[A. Ouyed, R. Ouyed, D, Leahy, and P. Jaikumar]{A. Ouyed$^{1}$, R. Ouyed$^{1}$, D. Leahy$^{1}$ and P. Jaikumar$^{2}$\\
$^1$ Department of Physics and Astronomy, University of Calgary,
2500 University Drive NW, Calgary, Alberta, T2N 1N4 Canada\\
$^2$ Department of Physics and Astronomy, California State University  Long Beach,  1250 Bellflower Blvd., Long Beach CA 90840}
\begin{document}

\pagerange{\pageref{firstpage}--\pageref{lastpage}} \pubyear{2002}

\maketitle

\label{firstpage}

\begin{abstract}
Current cc-SN models predict overproduction of $^{44}$Ti compared to observations. We present a model for an alternative channel where a cc-SN explosion is followed by a neutron star detonation (Quark Nova or QN), resulting in a spallation reaction of SN ejecta that produces $^{44}$Ti.  We can achieve a  $^{44}$Ti production of $\sim 10^{-4} \text{ M}_{\odot}$ with our model under the right time delay between the QN and the SN. Our model also produces unique signals not found in standard, cc-SN nucleosynthesis models. Some of these unique signals include a significantly large production of $^7$Be and $^{22}$Na. We discuss some of these signals by analyzing the late time light curve and gamma spectroscopy of our model.
\end{abstract}


\section{Introduction}

Late time light curves of SNe are considered a relatively accessible resource for  studying the nature of SNe, yet, the study of these light curves is ridden with challenges. Very few objects have been studied beyond $\sim 200$ days, because as the SN dims out, the dust and the background noise, makes the signals unreadable \citep{leibundgut2003optical}.  Even the most extensively documented supernova, SN1987A, is difficult to study at its later stages, because it is very hard to extrapolate directly the bolometric luminosity after $\sim1400$ days \citep{leibundgut2003optical}.  
The lack of late time light curve data make it difficult to formulate standard SN models.  SN models have their parameters adjusted so that the isotopic yields reflect the abundances inferred from late time light curves \citep{limongi2006nucleosynthesis}, which makes them dependent on the quality and understanding of our current observations. An example of the problematic nature of standard core collapse SN models, is $^{44}$Ti production. Although $^{44}$Ti production is assumed to be universal in core collapse SN models \citep{woosley1995presupernova}, only SN1987A is documented well enough to be able to extrapolate $^{44}$Ti from the bolometric light curve. The problem is precipitated by the fact that current $\gamma$-ray telescopes have only been able to detect $^{44}$Ti decay lines in Cassiopeia A \citep{the2006ti}, whereas current models predict more detections. \\
 	In this letter, we attempt to solve some  of these problems by proposing an alternative channel to the standard cc-SN model. After the SN explosion, neutron stars that undergo an evolution (e.g. matter fallback, or spin-down) evolution towards quark deconfinement densities will transition to a lower energy, strange matter phase and experience an energetic detonation.  (Quark Nova or QN) \citep{ouyed2002quark}. The neutron star will release its outer layers as relativistic, neutron rich ejecta, where the ejecta interacts with the SN envelope (double-shock Quark Nova, or dsQN). The time delay (hereafter $t_{\text{delay}}$) between the two detonations constrains the nature of the interaction, where a time delay on the order of days will cause significant spallation of SN ejecta by QN neutrons.  The spallation reaction creates unique signals in the late time light curve due to the modification of radioactive isotope abundances. The time delay $t_{\text{delay}}$ between the QN and SN detonations acts as a parameter for the isotopic abundances.  The different abundances of isotopes will also create specific gamma spectroscopic signals. Currently, there has been many indirect observations for QN suggested (e.g. \citet{hwang2012chandra,ouyed2012inprint,leahy2008supernova}).
\\
	In continuation to \citet{ouyed2011}, we explain the $^{44}$Ti synthesis measured in the late-time light curves of core collapse SN \citep{woosley1991co} as a product of nuclear spallation from the dsQN, in contrast to the traditional view that it is produced by the SN in situ.  The conditions necessary for the production of a QN, and the constraints set by the parameter $t_{\text{delay}}$, explains the rarity of observations of $^{44}$Ti nuclear decay lines. In contrast to our earlier paper \citep{ouyed2011}, we avoid excessive destruction of $^{56}$Ni by spallation, by taking into account the mixing of the SN's layers. Numerical studies have shown that Rayleigh Taylor instabilities can cause considerable mixing in a very short time span \citep{kifonidis2000nucleosynthesis,hachisu1991rayleigh,mueller1991instability}. Furthermore, observations in 1987A seem to support the picture of thorough mixing \citep{mueller1991instability}. If $^{56}$Ni is mixed through the SN envelope, there are less $^{56}$Ni atoms in the inner layer. This would lead to QN neutrons hitting less $^{56}$Ni, as opposed to the picture where all $^{56}$Ni is concentrated in the inner layers \citep{ouyed2011}. By avoiding excessive nickel depletion, we can avoid the sub-luminosity characteristic in \citet{ouyed2011}. 
	
\section{Model}\label{model}
	We extend the model we presented in \citet{ouyed2011} by including mixing. In analogy to lab terminology, we divide our model into a beam part, and a target part.  The scenario consists of a beam of relativistic QN ejecta that collides with a target that is the expanding, inner layers of the SN envelope. The time delay between the SN denotation and neutron star detonation is the key parameter in our model. 
	
\subsection{Beam}
	According to past studies \citep{keranen2005neutrino,ouyed2005fireballs},  a detonating neutron star can eject  its outer most layers as a mass of $M_{QN}\sim 10^{-3} \text{ M}_{\odot}$. We can model the QN ejecta as a pulse of $N^{0} \sim 1.2 \times 10^{54} M_{QN,-3}$ neutrons where $M_{QN,-3}$ means $M_{QN}$ in units of $10^{-3} \text{ M}_{\odot}$. The neutrons are relativistic, with an energy of $\sim 10$ GeV \citep{ouyed2011}.  
\subsection{Target}
	We model the target as an expanding SN envelope of inner radius $R = v_{\text{SN}} t_{\text{delay}}$, where $v_{\text{SN}} = 5000$ km/s is the velocity of the SN ejecta,  and $t_{\text{delay}}$ is the time delay between the neutron star detonation and the SN detonation. We assume the exploding star is $\sim 20 \text{ M}_{\odot}$. In contrast to the original model  \citep{ouyed2011}, we assume mixing. The target is the inner $ M_{\text{inner}} \sim1 \text{ M}_{\odot}$  mass of the expanding SN envelope, which is composed of mixed $^{16}$O,$^{12}$C, and $^{56}$Ni. We assume that whole SN envelope has a total $^{56}$Ni mass of $\sim0.1 \text{ M}_{\odot}$, and O and C have total masses of  $\sim 1.0 \text{ M}_{\odot}$ each \citep{ouyed2012}. If the  mixing has spread each element homogeneously through whole SN envelope, then we can derive the masses of each element for the interior $\sim 1.0 \text{ M}_{\odot}$ target. The masses of the isotopes in the target are $M_{\text{Ni}}=0.0476  \text{ M}_{\odot}$, and $M_O=M_C=0.476  \text{ M}_{\odot}$.  
	We treat the target, total number density $n_{\text{total}}$ as constant, where,
{\begin{equation} \label{dens}n_{\text{Total}} = n_{\text{Ni}}+ n_{\text{O}}+ n_{\text{C}} =\frac{N_{\text{total}}}{4 \pi R^2 \Delta R},  \end{equation}

where $\Delta R$ is the thickness containing $M_{\text{inner}}$ and $N_{\text{total}}$ is total number of atoms within $\Delta R$. Our mean free path $\lambda$ is,
\begin{equation}\label{lam} \lambda=\frac{1}{n_{\text{Ni}}\sigma_{56}+n_{\text{C}}\sigma_{12}+n_{\text{O}}\sigma_{16}},\end{equation}

where we use the semi-empirical cross section $\sigma_A=45\text{ mb}\times A^{0.7}$ \citep{letaw1983proton}, where $A$ is mass number. For our model, we use \eqref{dens} and \eqref{lam} to divide the thickness $\Delta R$ into $N_{\text{coll}}$ imaginary layers of width $\lambda$,

\begin{equation}N_{\text{coll}}=\frac{\Delta R}{\lambda} \approx 57.54 \frac{\frac{M_{\text{Ni},0.1 \text{ M}_{\odot}}}{56^{0.3}}+\frac{M_{\text{O},0.1 \text{ M}_{\odot}}}{16^{0.3}}+\frac{M_{\text{C},0.1 \text{ M}_{\odot}}}{12^{0.3}}}{v_{SN,1000 \text{km/s}} \times t_{\text{ delay},10\text{ days}}}, \end{equation}
where $N_{\text{coll}}$ represents the number of collisions an incoming neutron with $\Delta R$. The subscript indicates the units, for example, $M_{\text{O},0.1 \text{ M}_{\odot}}$ is in units of $0.1 \text{ M}_{\odot}$. The interaction of that neutron with a SN atom will release a multiplicity $\zeta$ of nucleons,

\begin{equation} \overline{\zeta}(E,A) \approx 4.67A_{56}(1+0.38\text{ln}E) Y_{\text{np}},\end{equation} where

 $1.25<Y_{\text{np}}<1.67$ \citep{cugnon1997nucleon,ouyed2011}. 
 
 \subsection{Spallation Statistics} 

At each imaginary layer $k$ of radial thickness $\lambda$, an atom will be bombarded by $N^0_{\rm{hits}}$ neutrons, resulting in the nucleus \citep{ouyed2011},

\begin{equation} A^1=A^0-\sum_{j=0}^{N^0_{\text{hits}}-1} \zeta^0(E^0,A^j). \end{equation}	
Each element in the layer has a statistical weight $w_i \sim N_i/N_{\text{total}} \sim M_{i}/(N_{\text{total}} A_{i} m_{H})$, where $N_i$ is the total number of atoms of isotope $i$, $M_{i}$ is the inner mass of a specific isotope $i$, $m_H$ is the mass of a proton, and $A_{i}$ is the atomic number of isotope $i$. The statistical weight signifies the fraction of atoms  of that particular element that will be hit by the incoming neutrons.

$N^0_{\rm{hits}}$ and $\zeta_0$ are drawn from a Poisson distribution that peaks at,

\begin{equation} \overline{N}^0_{\text{hits}}\sim \left ( \frac{N_{\text{Ni}}\sigma_{\text{Ni}}+N_{\text{C}} \sigma_{\text{C}}+N_{\text{O}}\sigma_{\text{O}}}{N_{\text{Ni}}+N_{\text{C}}+N_{\text{O}}} \right) \left(\frac{N^0 (1-e^{-1})}{4 \pi R^2}\right),\end{equation}

where $N_i$ is the number of atoms of isotope $i$, and $\sigma_i$ is the spallation cross section for isotope $i$.

For subsequent layers, it follows that.

 \begin{equation}\overline{N}^k_{\text{hits}}=(1-e^{-1})\overline{\zeta}^{k}\overline{N}^{k-1}_{\text{hits}} \end{equation}
 
and the mean nucleon energy for each subsequent layer is,

\begin{equation} E^ k\sim \frac{E^{k-1}}{\overline{\zeta}^{k-1}} \end{equation}

where spallation ceases at an energy of $E \sim 73$ MeV \citep {cugnon1997nucleon}.

\subsection{Light Curve Model}

We compute the late time light curve by extending our spallation code with the Cococubed program (Timmes). Our spallation code computes the mass yields of different isotopes using the model in Section \ref{model}, and the Cococubed extension computes the bolometric luminosity from the isotope mass yields.  We also take into account the contributions of  $^{57}$Ni, and $^{60}$Co to the bolometric light curve, and we take their mass values from \citet{woosley1995presupernova}.  In our model, we assume the simple case where the bolometric luminosity is entirely caused by the thermalization of nuclear decay radiation.

	\section{Results}

\subsection{Photometry}

The key isotopes for the light curve (Fig. \ref{lightcurves}) are $^{56}$Ni and $^{44}$Ti.  For $t_{\text{delay}} = 4$ days, our model synthesized  $\sim 10^{-4}  \text{ M}_{\odot}$ of $^{44}$Ti, which compares favourably to amounts found by $\gamma$-ray telescopes in Cassiopeia A \citep{iyudin1994comptel}, and amounts estimated from the light curve of SN1987A \citep{suntzeff1992energy,woosley1991co}. Furthermore, we notice there is very little nickel that is depleted - indeed most of the original inner $0.0476  \text{ M}_{\odot}$ is maintained for $3 \text{ days}\leq t_{\text{delay}}\leq 10 \text{ days}$. If very little $^{56}$Ni is depleted from the inner mass, then the total $\sim 0.1 \text{ M}_{\odot}$ for $^{56}$Ni that is spread throughout the whole envelope, is more or less maintained throughout different time delays. The  $\sim 0.1 \text{ M}_{\odot}$ mass  for $^{56}$Ni compares quite well with what is observed in type II SN light curves \citep{suntzeff1992energy,woosley1991co}. Another interesting isotope is $^{22}$Na, and it is synthesized at a high amount of almost $\sim 10^{-4} \text{ M}_{\odot} $ which contrasts with the estimates of $\sim 10^{-6}\text{ M}_{\odot}$ \citep{woosley1995presupernova}. This creates a higher contribution to the bolometric luminosity, than normally expected.
	We see in the upper panel of Fig \ref{lightcurves} a  dependance of the light curve to $t_{\text{delay}}$. We can see that latter stages of the late time light curve are the most visibly affected by the time delay. This strong dependance in the later days is modulated by the production of $^{44}$Ti, because $^{44}$Ti yields are very affected  by the time delay (Fig. \ref{massn2}). The earlier days of the late time light curve seem to be affected only weakly by the time delay change, because $^{56}$Ni destruction stays more or less constant (Fig. \ref{massn2}).
   
\begin{figure}
 \centering
    \includegraphics[width=0.5\textwidth]{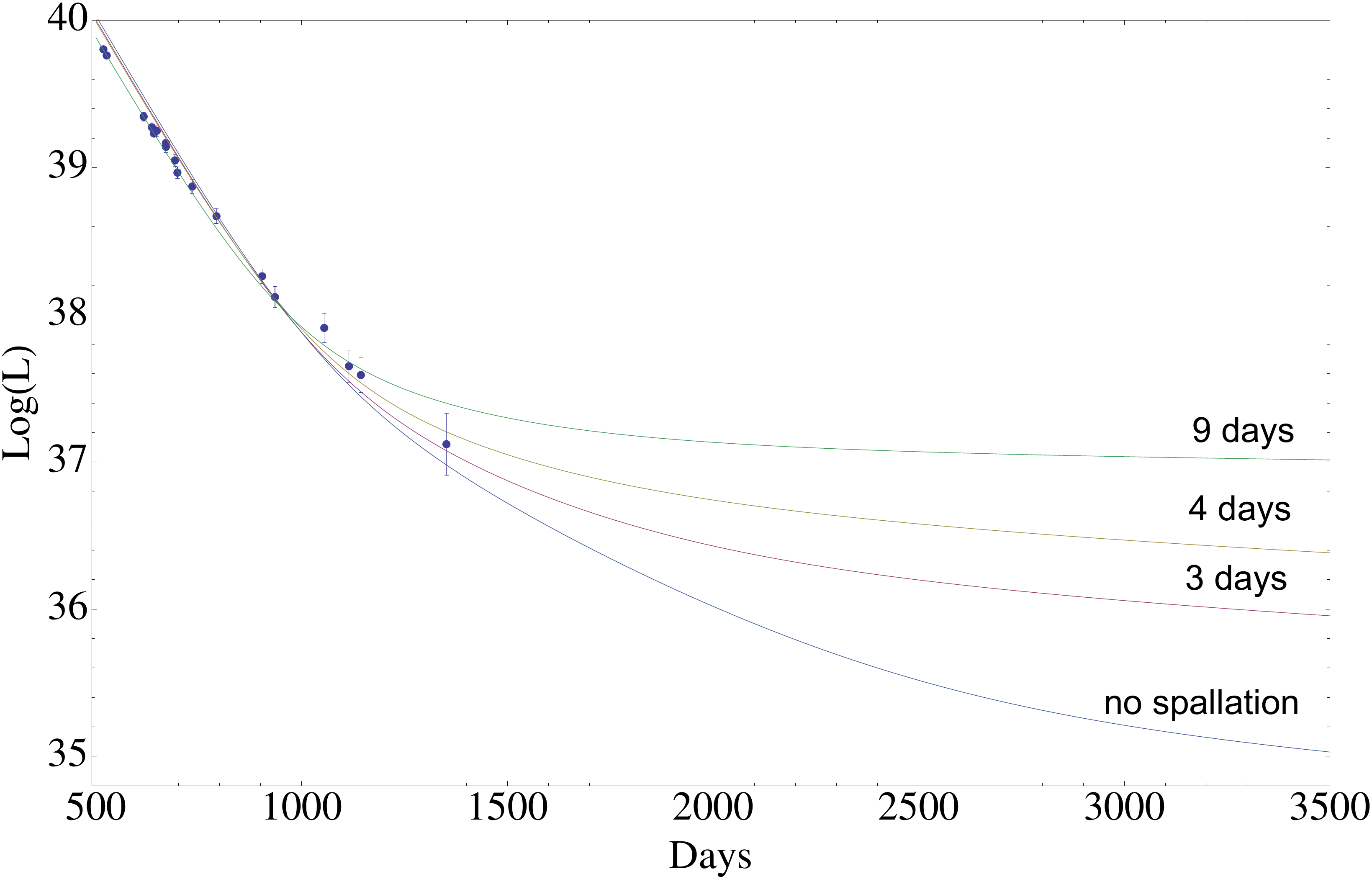}
    \includegraphics[width=0.5\textwidth]{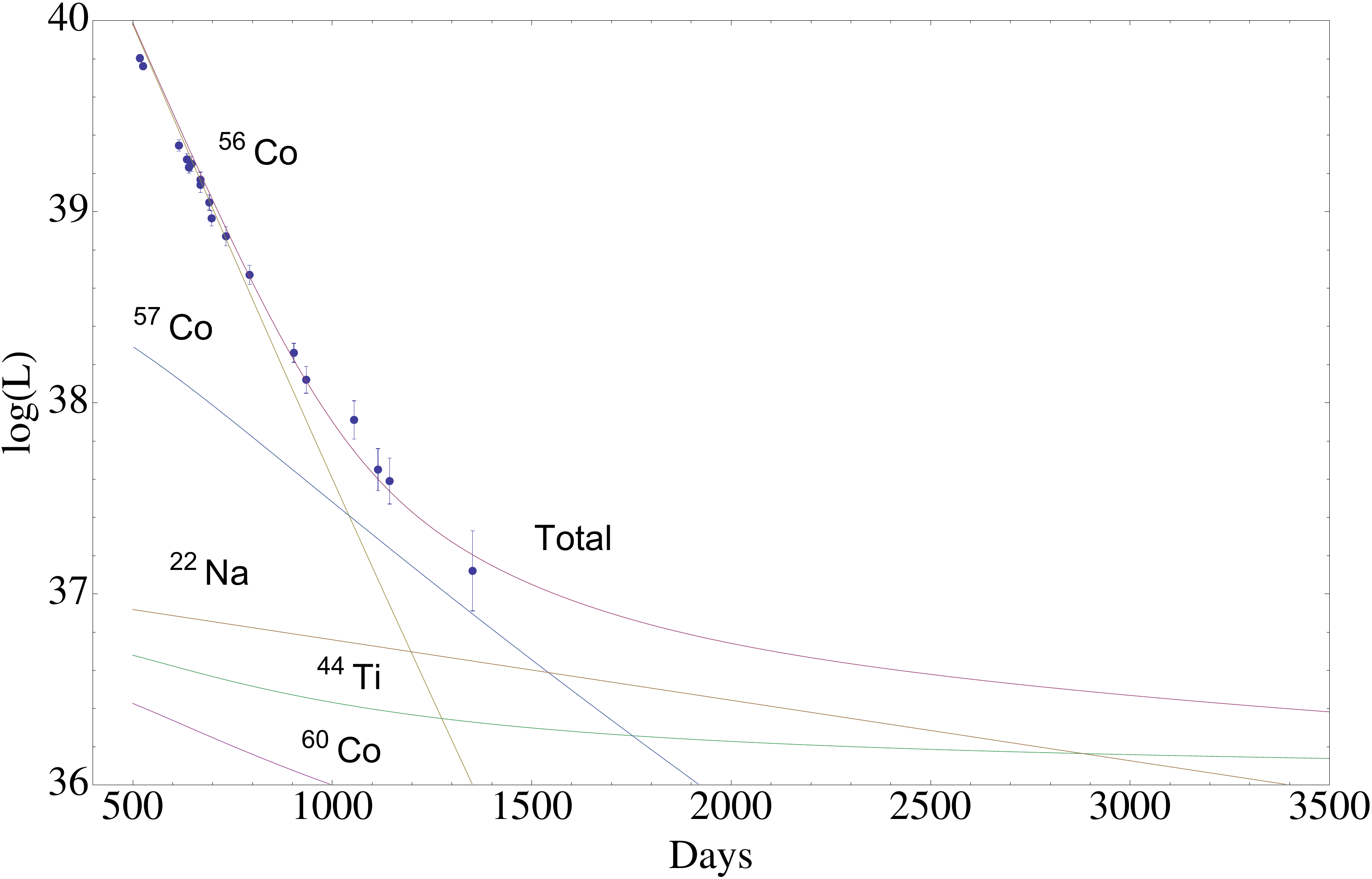}
    
  \caption{ \label{lightcurves}
The upper  panel represents the light curve data of SN1987A  against our model for different time delays.Ê The lower panel depicts the SN1987A data with the   $t_{\text{delay}}=4$  model.  The SN1987A data comes from \citet{suntzeff1991late}. The time delays represented in the upper panel are $t_{\text{delay}}=3,4$ and $9$ days. In the upper panel we also depict the original $M_{\text{Ni}}=0.1 \text{ M}_{\odot}$ target without the spallation, where luminosity contributions of  $^{57}$Co and $^{60}$Co were added artificially, and their values, $5.5\times10^{-3} \text{ M}_{\odot}$ and  $1.14\times10^{-5} \text{ M}_{\odot}$, respectively, were taken from \citet{woosley1995presupernova}.  The points are the SN 1987A data and the solid lines indicate our model. The y-axis is the logarithm of luminosity in units of erg/sec. In the lower plot, we also include the individual luminosity contributions of each isotope, where the contributions of $^{57}$Co and $^{60}$Co were artificially constructed in identical fashion to the upper panel. The masses of $^{56}$Ni, $^{22}$Na and $^{44}$Ti,  are actual, spallation products of our model.  For $t_{\text{delay}}=4$ days, the values are $0.99  \text{ M}_{\odot} $, $ 5.50 \times 10^{-5} \text{ M}_{\odot}$, and $1.20\times 10^{-4} \text{ M}_{\odot} $, respectively. }
 
\end{figure}

\subsection{Isotope Abundance} 
	 
\begin{figure}
 \centering
    \includegraphics[width=0.44\textwidth]{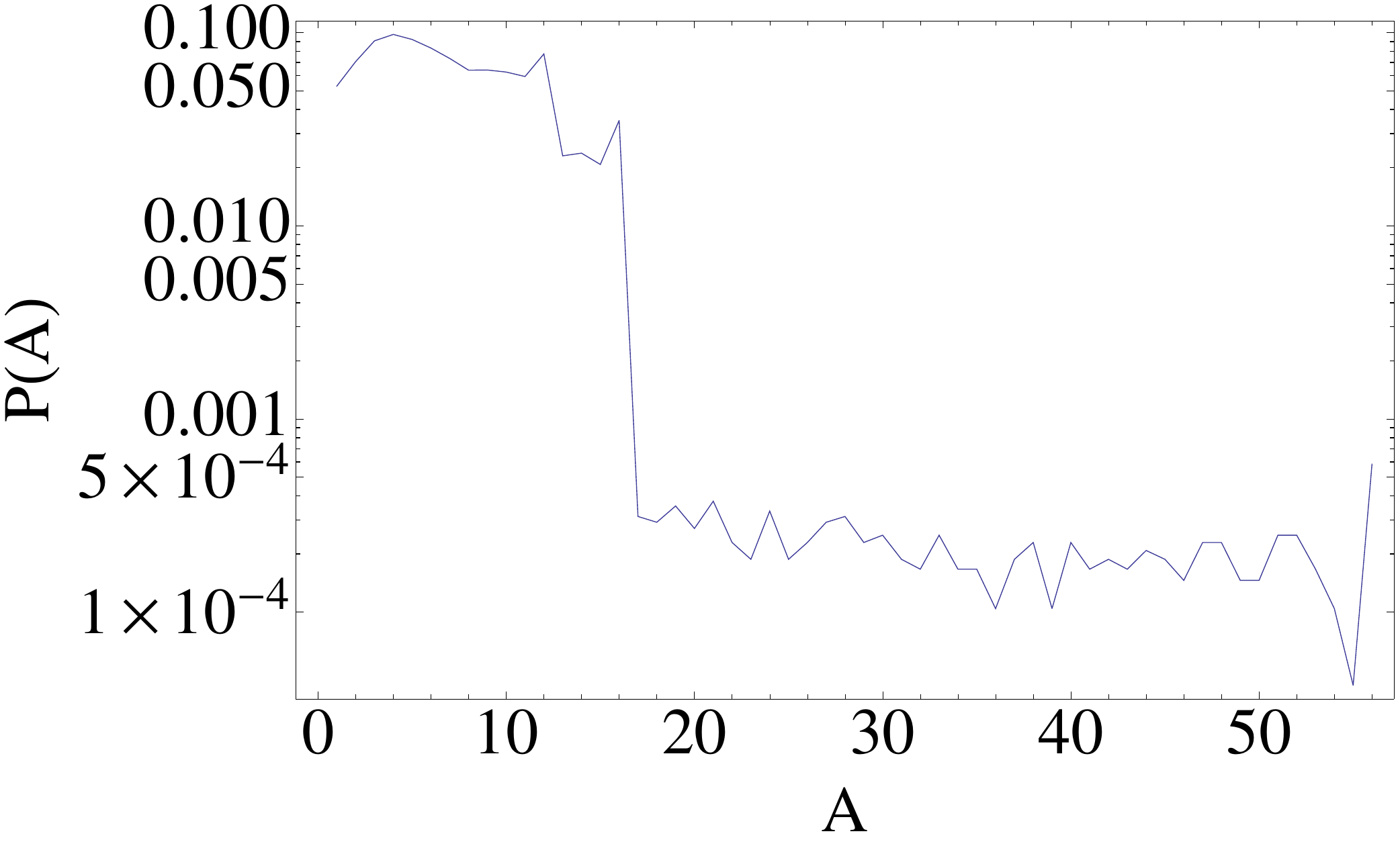}
           \caption{ \label{massn1} The  panel represents the fractional abundance of atoms of a particular atomic mass number produced by  our spallation model at a $t_{\text{delay}}= 4$ days, where the x-axis represents the mass number A, and the y-axis represents the fractional probability of nucleus of mass number A. The panel depicts the yields for spallation of the inner $\sim 1.0  \text{ M}_{\odot}$ of the SN ejecta.}   
 
\end{figure}

\begin{figure}
 \centering
 \includegraphics[width=0.44\textwidth]{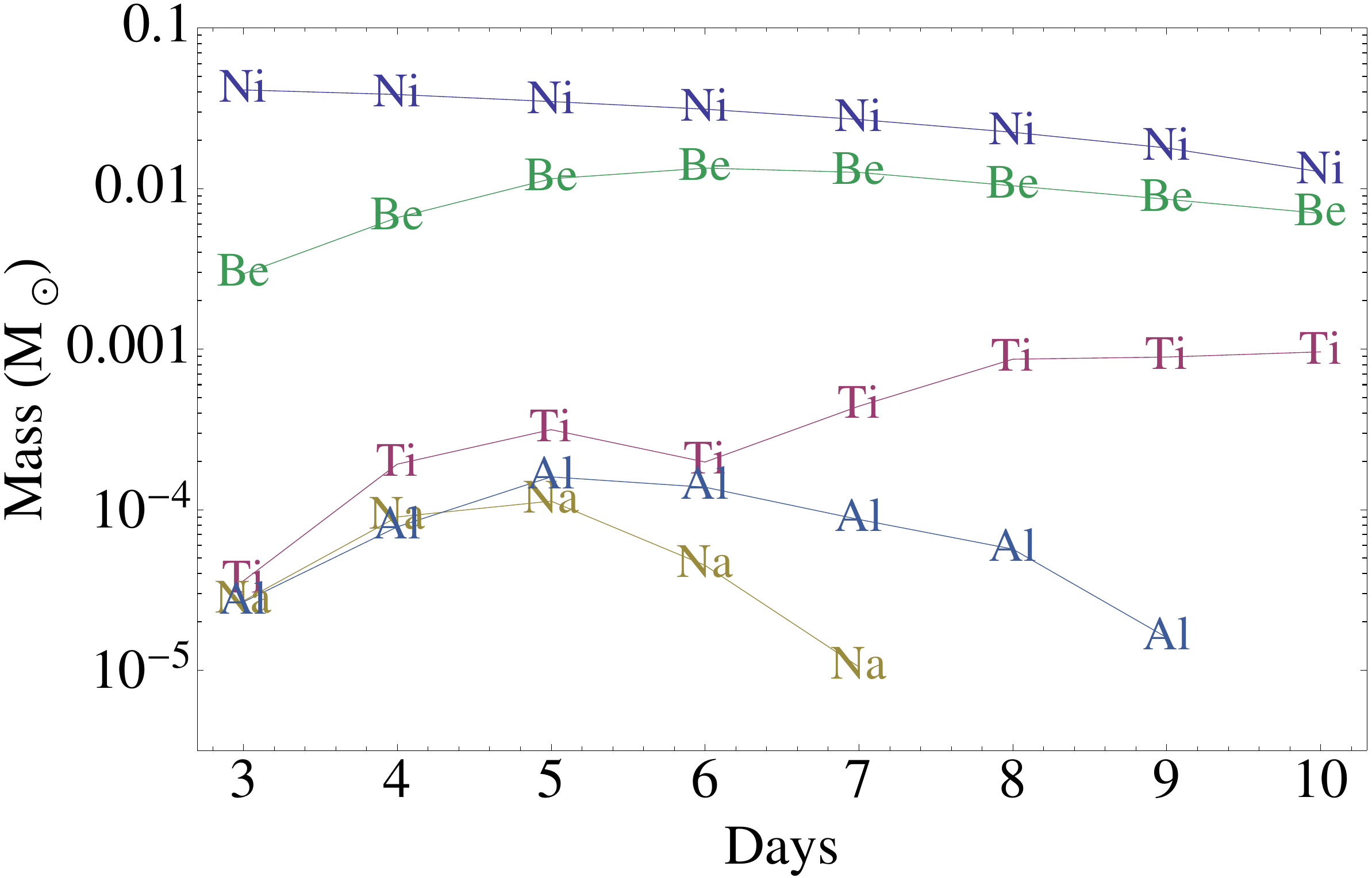}
       \caption{ \label{massn2} The  panel represents the mass yields of  $^{56}$Ni, $^{44}$ Ti, $^{22}$Na, $^{26}$Al and $^7$Be, where x-axis represents $t_{\text{delay}}$ and the y-axis represents the mass in units of $\text{ M}_{\odot}$. The panel depicts the yields for spallation of the inner $\sim 1.0  \text{ M}_{\odot}$ of the SN ejecta.}   
 
\end{figure}

	Our spallation model creates isotopes of  almost every mass number in the range of $A<56$.  However, very few of the isotopes created can be detected with $\gamma$-ray telescopes, because the SN envelope only becomes transparent to nuclear decay radiation after a time period of at least months \citep{diehl2012nucleosynthesis}, so only isotopes that have a half life of at least in the order of weeks will have detectable nuclear decay lines. Therefore, in Fig \ref{massn2}, we only include the masses of isotopes with half lives of at least in the order of weeks. 
	
	In Fig. \ref{massn1} and  \ref{massn2}   we plot some of the trends of these isotopes. We notice that in  Fig. \ref{massn1}, the plot heavily leans on light isotopes. This is caused by the spallation of mixed C and O in the inner layer. We also notice that the plot in  Fig. \ref{massn1} is divided in roughly two plateaus, where their boundary is in $A\sim16$, which is the atomic mass number of Oxygen.  In Fig. \ref{massn2} we plot the mass of one of the lighter isotopes, $^{7}$Be, which has a half life of 76 days, which gives enough time for nuclear decay photons to go through the SN photosphere.  Compared to the models of \citet{woosley1995presupernova}, we notice that our model produces more $^{7}$Be and $^{22}$Na. We also notice that $^{44}$Ti, $^{56}$Ni, and $^{26}$Al are in the same order of magnitudes as  other models \citep{woosley1995presupernova}. Finally, it seems we can constrain production of $^{22}$Na  and $^{26}$Al at higher $t_{\text{delay}}$, where $^{22}$Na  production ends at $t_{\text{delay}}\sim 7$ and $^{26}$Al production at $t_{\text{delay}} \sim 8$ days. 
	
\section{Discussion and Predictions}

\begin{itemize}
\item {\it $^{44}$Ti production as a dsQN specificity and SN1987A}
We argue that the rarity of observations \citep{the2006ti} of $^{44}$Ti might be related to the fact that $^{44}$Ti is not produced by standard cc-SNe, but by dsQNe. In a recent paper \citep{ouyed2011}, we argued that CasA's $^{44}$Ti lines were produced by a dsQN through a significant $^{56}$Ni depletion. Although, it is probably possible to avoid the excessive $^{56}$Ni depletion in our original model by adjusting some parameters, and yet still predict a $^{44}$Ti mass yield that is compatible with observations, it seems more natural and elegant to extend our model with mixing.  Through mixing, we can predict very minimal destruction of $^{56}$Ni (Fig. \ref{massn2}) because $^{56}$Ni is spread throughout the whole envelope as opposed to the assumption where it is all concentrated in the inner layers. If $^{56}$Ni is spread throughout a wider volume, then less $^{56}$Ni atoms will be available in the inner layers as targets for QN neutrons.  This minimal depletion leaves almost all the  $\sim 0.1 \text{ M}_{\odot}$ of $^{56}$Ni intact.  In Fig \ref{lightcurves} we plotted the data of SN1987A  against our model, where SN1987A is a SN where $^{44}$Ti has been measured indirectly \citep{motizuki2004radioactivity,suntzeff1992energy,woosley1991co}. At lower time delays, QN energy is mostly spent in PdV work \citep{leahy2008supernova}, therefore at a time delay of 4 days, our model could reproduce a similar luminosity to SN1987A .  Current observations of SN1987A point to thorough mixing of SN layers \citep{mueller1991instability}, which compares well to the assumptions about mixing done in our model. Recent papers have provided alternative arguments for the possibility of SN1987A compact remnant being a Quark Star \citep{chan2009could}.  

\item {\it $^7$Be abundance }
	In Fig. \ref{massn1} and \ref{massn2}, a very large abundance of $^{7}$Be is produced compared to current cc-SN models \citep{diehl1998gamma,woosley1995presupernova}. Current core collapse SN models don't produce enough $^7$Be to reach the photosphere before $^{7}$Be decays. However, our model predicts $\sim 10^{-3}$ - $10^{-2} \text{ M}_{\odot}$ for $^7$Be, which could be enough so that a significant amount of it reaches the photosphere and therefore be detected by $\gamma$-ray telescopes.  The 487 keV  photon released by the $\beta$-decay of $^7$Be into $^7$Li, is in the detectable energy range of current $\gamma$-ray telescopes (INTEGRAL, NuSTAR, etc.).  
	The large abundance of $^{7}$Be is due to the spallation of mixed O and C in the inner SN layers. This high abundance in our model indicates that only core collapse SN that evolve into  dsQN will have a detectable gamma signature of $^7$Be.  Currently, Novae are the only possible sources of detectable, $^{7}$Be photons because standard cc-SN are too optically thick \citep{diehl1998gamma}. However our dsQN model can produce an alternative point source of $^{7}$Be photons. 
	Finally, such a massive amount of $^{7}$Be should create a signature in the bolometric light curve.
	
	\item {\it $^{22}$Na abundance}	
	A small  $^{22}$Na  contribution is predicted in late time core collapse SN light curves \citep{woosley1995presupernova}. However, our spallation model generates a much larger amount of  $^{22}$Na  than expected. This large yields of $^{22}$Na  are due to the spallation of $^{56}$Ni into lighter nuclei. Current models \citep{woosley1995presupernova} give a $\sim 10^{-6} \text{M}_{\odot}$ for $^{22}$Na, while our spallation model can produce as much as $ \sim 10^{-4} \text{M}_{\odot}$ (Fig \ref{massn2}). Furthermore, $^{22}$Na has a  half-life of 2.6 yr, and decays into an excited state of $^{22}$Ne, releasing energetic radiation \citep{diehl1998gamma}. That large amount generates enough radiation to create a noticeable signal in the late time light curve, as pictured in Fig \ref{lightcurves}.  The decay of $^{22}$Na  emits a 1.275 MeV $\gamma$ line that can be detected by current $\gamma$-ray telescopes. These  signals of $^{22}$Na  abundant dsQN cannot be reproduced in current models for cc-SNe. 
		
	\item {\it Mixing}
	Our model makes the simplification that O,C and $^{56}$Ni are  in a perfect, homogenous mix. Our simplification allows us to model the spallation targets with statistical weights derived from the isotopic abundances. However, Rayleigh Taylor instabilities induce mixing by  fast moving, mushroom-like structures, structures that are lost when assuming a perfect, homogenous mixture.  Furthermore, studies of CasA's remnants point to considerable asymmetry in the detonation \citep{fesen2006expansion}, which seems to contradict the assumption of perfect homogenous mixing. However, the main ideas of our model, that $^{44}$Ti can be produced from a minimal destruction of $^{56}$Ni, that spallation of $^{56}$Ni will lead to enrichment of $^{22}$Na, and that spallation of O, and C will lead to an excess of $^7$Be, seem sound. While other limitations of our model exist, which are discussed in \citet{ouyed2011}, we offer some sound predictions for $^{22}$Na and $^7$Be enrichment.

\end{itemize}

\section*{Acknowledgments}
This research is supported by an operating grant from the National Science and Engineer- ing Research Council of Canada (NSERC). We also thank N. Koning and M. Kostka for helpful discussion.

\bibliographystyle{mn2e} 
\bibliography{snbib}

\label{lastpage}

\end{document}